%% file: eps09_hoecker.tex
\documentclass{PoS}

\usepackage{refmerge}
\usepackage{xspace}

\include{definitions}

\title{Status of the global electroweak fit of the Standard Model}

\ShortTitle{Status of the global electroweak fit of the Standard Model}

\author{\speaker{Andreas Hoecker}\thanks{For the Gfitter Group: 
   M.~Baak (CERN), M.~Goebel (DESY and U. Hamburg), H.~Fl\"acher (CERN), 
   J.~Haller (U. Hamburg), A.~Hoecker (CERN), D.~Ludwig (DESY and U. Hamburg), 
   K.~M\"onig (DESY), M.~Schott (CERN) and J.~Stelzer (DESY).}\\
  CERN CH-1211, Geneva 23, Switzerland \\ 
  E-mail: \email{andreas.hoecker@cern.ch}}

\abstract{
Results from the global Standard Model fit to electroweak precision data, including 
newest Tevatron measurements, are reviewed and discussed. The complete fit using also 
the constraints from the direct Higgs boson searches yields an upper limit on the Higgs 
mass of 153\gev at 95\% CL. The top mass is indirectly determined to be 
$177.2^{+10.5}_{-7.8}\gev$ and $179.5^{+8.8}_{-5.2}\gev$ for fits including or not
the constraints from the direct Higgs searches, respectively. Using the 3NLO perturbative
prediction of the massless QCD Adler function, the strong coupling constant at the 
$Z$-mass scale is determined to be $\asZ=0.1193\pm0.0028\pm0.0001$, which is in 
excellent agreement with the 3NLO result from hadronic $\tau$ decays. The perspectives 
of the electroweak fit for forthcoming and proposed future collider projects are discussed. 
The available constraints on the Higgs mass are convolved with the high-scale behaviour 
of the Higgs quartic coupling to derive likelihoods for the survival of the Standard 
Model versus its cut-off scale evolved up to the Planck mass.
}

\FullConference{The 2009 Europhysics Conference on High Energy Physics\\
		 Krakow, Poland\\
		 July 16 -- 22, 2009}

\begin{document}

\section{Introduction}
\label{introduction}

Precision measurements allow us, by exploiting contributions from quantum loops, to 
probe physics at much higher energy scales than the masses of the particles directly 
involved in experimental reactions. These tests do not only require accurate and well 
understood experimental data but also theoretical predictions with controlled 
uncertainties that match the experimental precision. Prominent examples are the 
LEP precision measurements, which were used in conjunction with the Standard Model 
(SM) to predict via multidimensional parameter fits the mass of the top 
quark~\cite{Alexander}, prior to its observation at the 
Tevatron~\cite{Abachi:1995iq}.\footnote
{
   The importance of radiative corrections at the EW scale can be illustrated by 
   comparing the tree-level EW unification prediction of the $W$ mass,
   $M_{W}^{(0)2}=(M_{Z}^2/2)(1+\sqrt{1-\sqrt{8}\,\pi\alpha/\GF{M_{Z}^2}})=(79.964\pm0.005)\gev$, 
   with the world average measurement, $M_W=(80.399\pm0.023)\gev$, 
   exhibiting a $18.5\sigma$ discrepancy due to a $0.5\%$ contribution from loop effects.
   The dominant one-loop diagrams are bosonic and fermionic vacuum polarisation and 
   self energies, as well as top--$W$ corrections to the $Z\to b\bbar$ vertex. Their 
   dependence on the top and Higgs boson mass parameters are respectively quadratic and 
   logarithmic.
} 
Later, when combined with the measured top mass, the same approach led to the prediction 
of a light Higgs boson~\cite{:1994qa}. 

Several theoretical libraries within and beyond the SM have been developed in the past,
which allowed to constrain the unbound parameters of the SM~\cite{Arbuzov,Montagna}.
However, most of these programmes are relatively old, were implemented in outdated programming 
languages, and are difficult to maintain in line with the theoretical and experimental progress. 
It is unsatisfactory to rely on them during the forthcoming era of the Large Hadron Collider (LHC)
and the preparations for future linear collider projects. Improved measurements of 
important input observables are expected and new observables from discoveries may augment 
the available constraints. None of the previous programmes are modular 
enough to easily allow the theoretical predictions to be extended to models beyond the SM,
and they are usually tied to a particular minimisation package. 

These considerations led to the development of the generic fitting package {\em Gfitter}~\cite{gfitter}, 
designed to provide a framework for model testing in high-energy physics. Gfitter is implemented 
in C++ and relies on ROOT~\cite{root} functionality, XML and python. Theoretical models 
are inserted as plugin packages. Tools for the handling 
of the data, the fitting, and statistical analyses such as pseudo Monte Carlo sampling are provided 
by a core package, where theoretical errors, correlations, and inter-parameter dependencies 
are consistently dealt with. The use of dynamic parameter caching avoids the recalculation 
of unchanged results between fit steps, and thus significantly reduces the amount of computing 
time required for a fit. 

These proceedings review the current status of the global electroweak fit and discuss the 
prospectives of the fit for forthcoming and future collider projects. Following Ref.~\cite{fate}
we also review the evolution properties of the quartic coupling in the SM Higgs potential to 
high scales, and combine them with the available constraints on the Higgs boson mass. More 
detailed information on the latter topic has been presented at this conference~\cite{espinosa}. 
Reference~\cite{baak} reviews the status of beyond SM constraints with Gfitter.

\section{The global electroweak fit of the Standard Model }

The SM predictions for the electroweak precision observables measured by the LEP, SLC, and 
Tevatron experiments are fully implemented in Gfitter. State-of-the-art calculations 
have been used, and the results were thoroughly cross-checked against
ZFITTER~\cite{Arbuzov}. For the $W$ mass and the effective weak mixing angle, which exhibit 
the strongest constraints on the Higgs mass, the full second order corrections are 
available~\cite{Awramik}. Furthermore, corrections 
of order ${\cal O}(\alpha \as^2)$ and leading three-loop corrections in an expansion 
of the top-mass-squared ($\mt^2$) are included. 
The full three-loop corrections are known in the large $M_H$ limit,
however they turn out to be negligibly small \cite{Boughezal}.
The partial and total widths of the $Z$ are known
to leading order, while for the second order only the leading $\mt^2$ corrections are 
available~\cite{Bardin:1997xq}. Among the new developments included in the SM library is 
the fourth-order (3NLO) perturbative calculation of the massless QCD Adler 
function~\cite{Baikov:2008jh}, contributing to the vector and axial-vector radiator functions 
in the prediction of the $Z$ hadronic width (and other observables). It allows to fit the 
strong coupling constant with unique theoretical accuracy.

Among the experimental precision data used are the $Z$ mass, measured with relative 
precisions of $2 \cdot 10^{-5}$ at LEP, and the hadronic pole cross section and leptonic 
decay width ratio of the $Z$, both known to $9\cdot10^{-4}$ relative precision. The 
effective weak mixing angle $\sinleff$ is known to a relative precision of $7 \cdot 10^{-4}$ 
from the measurements of the left-right and forward-backward asymmetries for universal leptons 
and heavy quarks by the LEP and SLD experiments. The $W$ mass has been measured at LEP 
and the Tevatron to an overall relative precision of $3 \cdot 10^{-4}$. We include the new 
preliminary result reported by D0~\cite{d0Wmass} and combine it with the previous world 
average to the preliminary average $80.399\pm0.023$, taking into account correlations 
between systematic errors.\footnote
{
   Our $M_W$ average agrees with the recently published official Tevatron value~\cite{mwtev}.
}
 The top mass has been measured to $7\cdot10^{-3}$ relative 
precision at the Tevatron. We use the newest average $m_t=(173.1\pm0.6\pm1.1)\gev$~\cite{tevtop},
where the first error is statistical and the second systematic. Also required is the
knowledge of the electromagnetic coupling strengths at the $M_Z$ scale, which 
is modified with respect to the Thomson scattering limit due to energy-dependent photon 
vacuum polarisation contributions. It is known to a relative precision of $8 \cdot 10^{-3}$,
dominated by the uncertainty in the hadronic contribution from the five lightest quarks, 
$\Dalphahad$. Finally, the Fermi constant, parametrising the weak coupling strength, is 
known to $4\cdot10^{-5}$ relative precision.

\begin{figure}[p]
  \newcommand\thisSize{0.660\textwidth}
  \begin{center}
    \includegraphics[width=\thisSize]{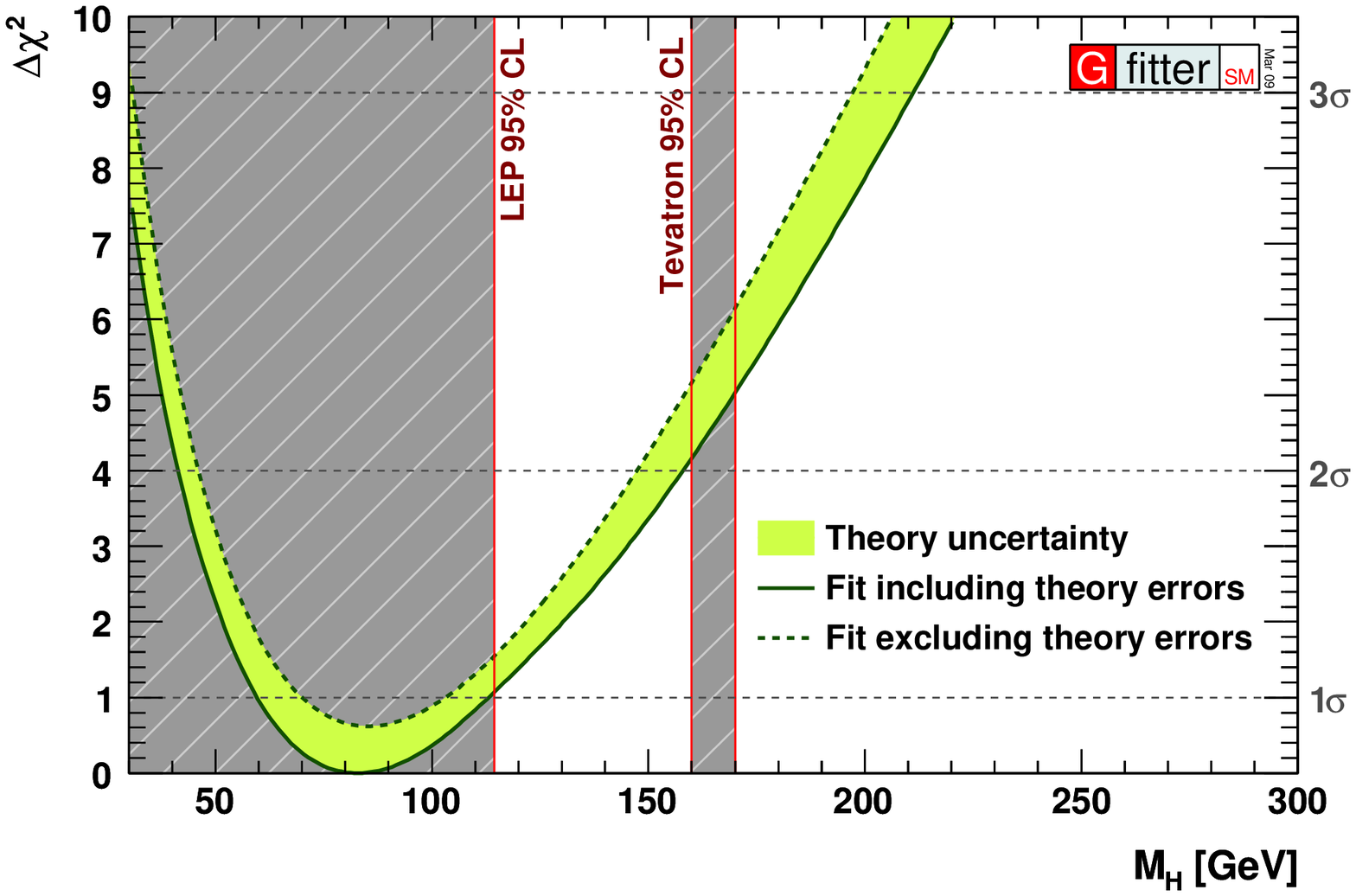}
    \includegraphics[width=\thisSize]{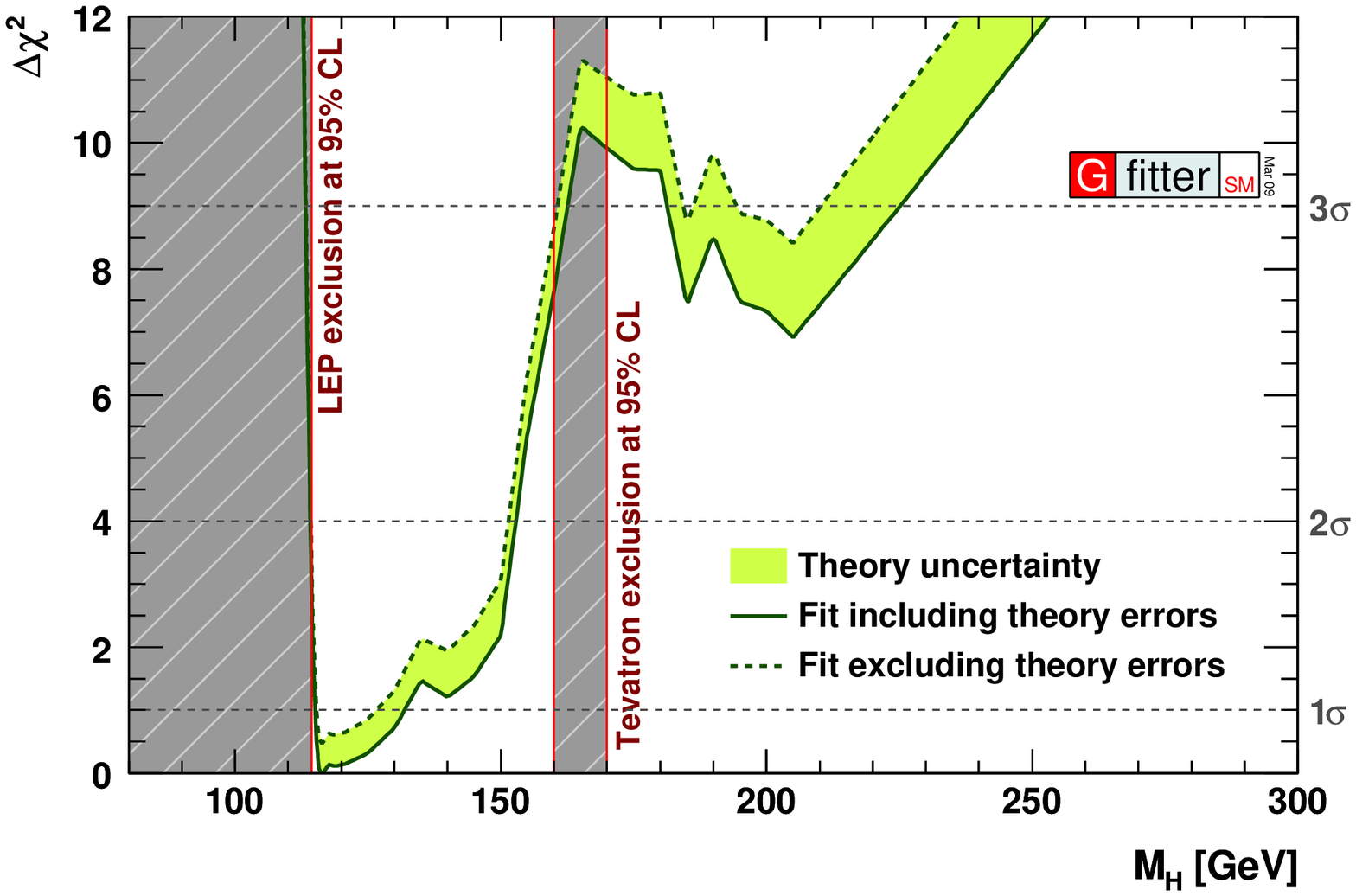}
    \includegraphics[width=\thisSize]{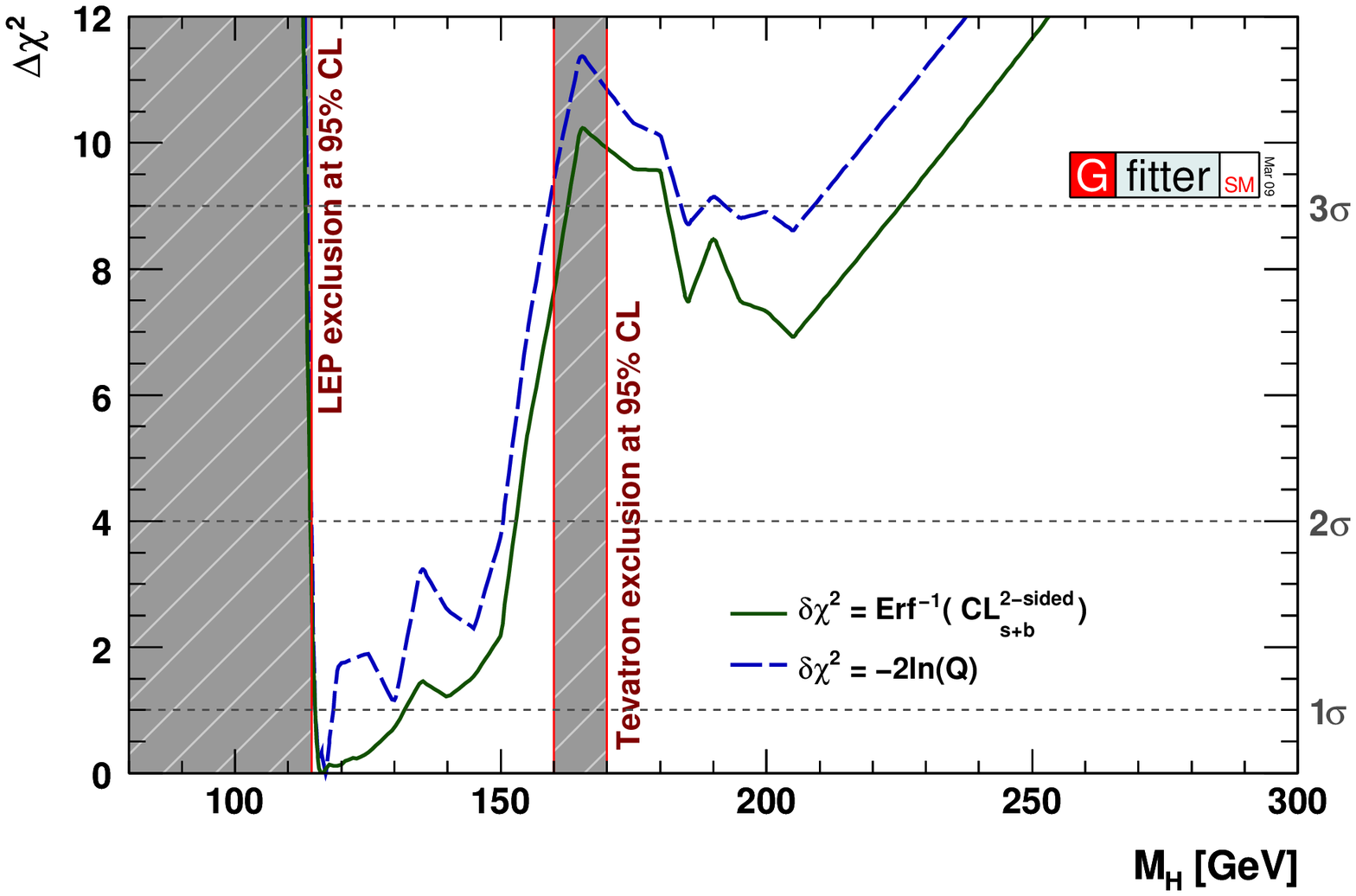}
  \end{center}
  \vspace{-0.3cm}
  \caption{\DeltaChi as a function of $M_H$ for the standard fit (top) and the 
            complete fit (middle). The solid (dashed) lines give the results when 
            including (ignoring) theoretical errors. The bottom plot shows the results from
            the complete fit with the two-sided $\CL_{\rm s+b}$ method (solid line, same as 
            middle plot) and the direct use of $\Delta\chi^2=-2\ln Q$ as estimator (dashed). 
            The minimum $\Delta\chi^2$ being deeper for the latter method, the overall 
            curve is shifted.}
\label{fig:mhiggs}
\end{figure}

We also fold into the fit the information from the direct Higgs boson searches at 
LEP~\cite{Higgs-LEP} and Tevatron, where for the latter experiments the latest 
combination is used~\cite{Higgs-Tev}, including a rising number of search channels with 
up to 4.2\invfb integrated luminosity. All experiments use as test statistics the 
negative logarithm of a likelihood ratio, $-2\ln\!Q$, of the SM Higgs signal plus 
background ($\rm s+b$) to the background-only ($\rm b$) hypotheses. This choice ensures 
$-2\ln\!Q=0$ when there is no experimental sensitivity to a Higgs signal. The 
corresponding one-sided confidence levels $\CL_{\rm s+b}$ and $\CL_{\rm b}$ describe 
the probabilities of upward fluctuations of the test statistics in presence and absence 
of a signal, respectively ($1-\CL_{\rm b}$ is thus the probability of a false discovery).
They are derived using pseudo Monte Carlo (MC) experiments. Using the modified quantity 
$\CL_{\rm s}=\CL_{\rm s+b}/\CL_{\rm b}$ the combination of LEP searches~\cite{Higgs-LEP} 
has set the lower limit $M_H>114.4\gev$ at 95\%~CL, and the Tevatron experiments
recently reported the exclusion of the range $160<M_H<170\gev$ at and above
95\%~CL~\cite{Higgs-Tev}. Because in the EW fit we are interested in the {\em deviation}
of a measurement from the SM hypothesis, we transform $\CL_{\rm s+b}$ into a two-sided
confidence level given by $2 \CL_{\rm s+b}$ for $\CL_{\rm s+b}\le0.5$ and $2(1-\CL_{\rm s+b})$ 
otherwise. The contribution to the $\chi^2$ estimator of the fit is then obtained via
$\delta\chi^2=2\cdot[{\rm Erf}^{-1}(1-\CL^{\rm 2\mbox{-}sided}_{\rm s+b})]^2$.
The alternative (Bayesian) direct use of the test 
statistics $-2\ln\!Q$ in the fit leads to a similar behaviour as $\delta\chi^2$
with however an overall shift of approximately one unit due to a deeper minimum,
thus resulting in a slightly stronger constraint on $M_H$ (\cf lower plot 
in Fig.~\ref{fig:mhiggs} with the $-2\ln\!Q$ curve drawn dashed).

\begin{figure}[p]
  \newcommand\thisSize{0.660\textwidth}
  \begin{center}
    \includegraphics[width=\thisSize]{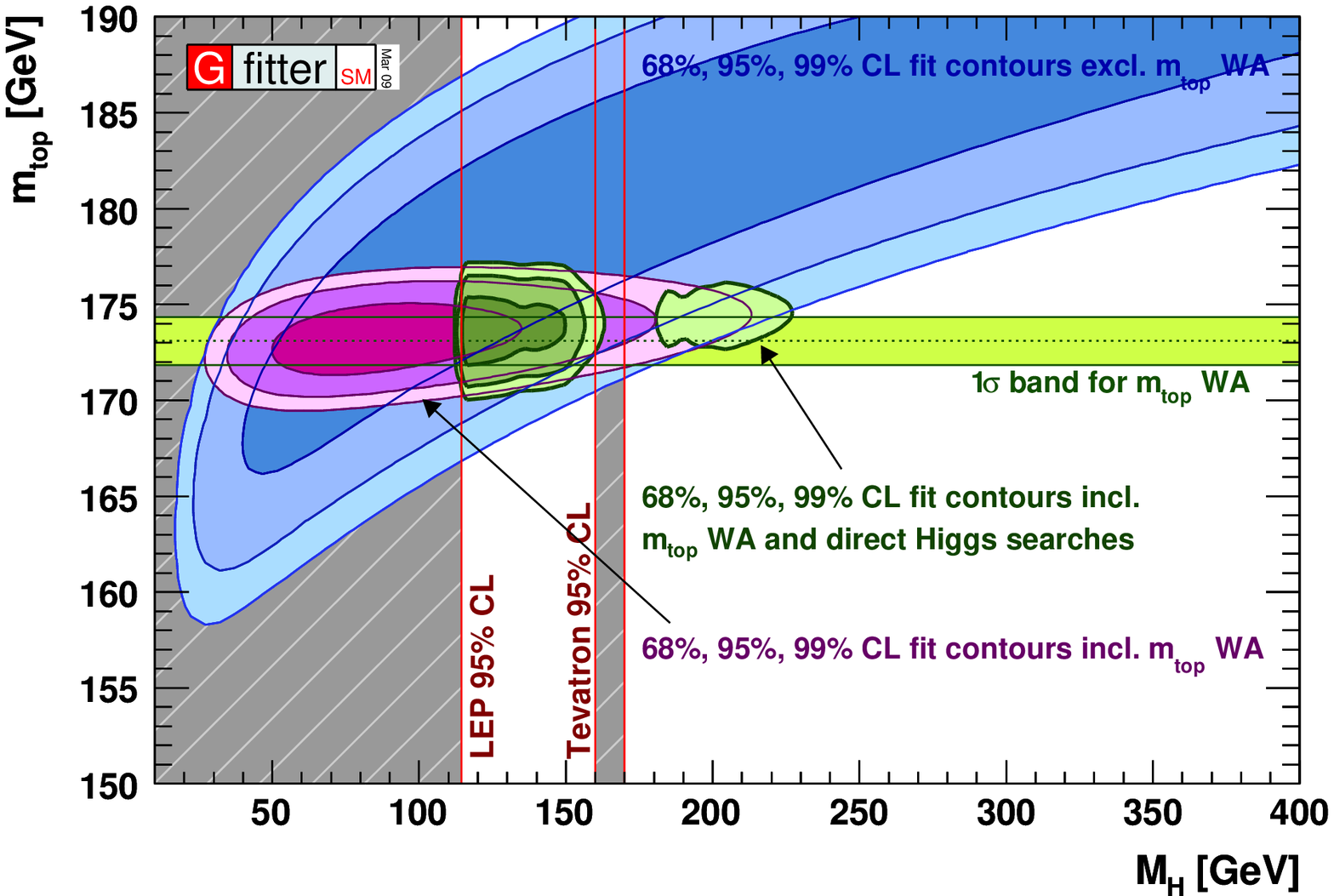}
    \includegraphics[width=\thisSize]{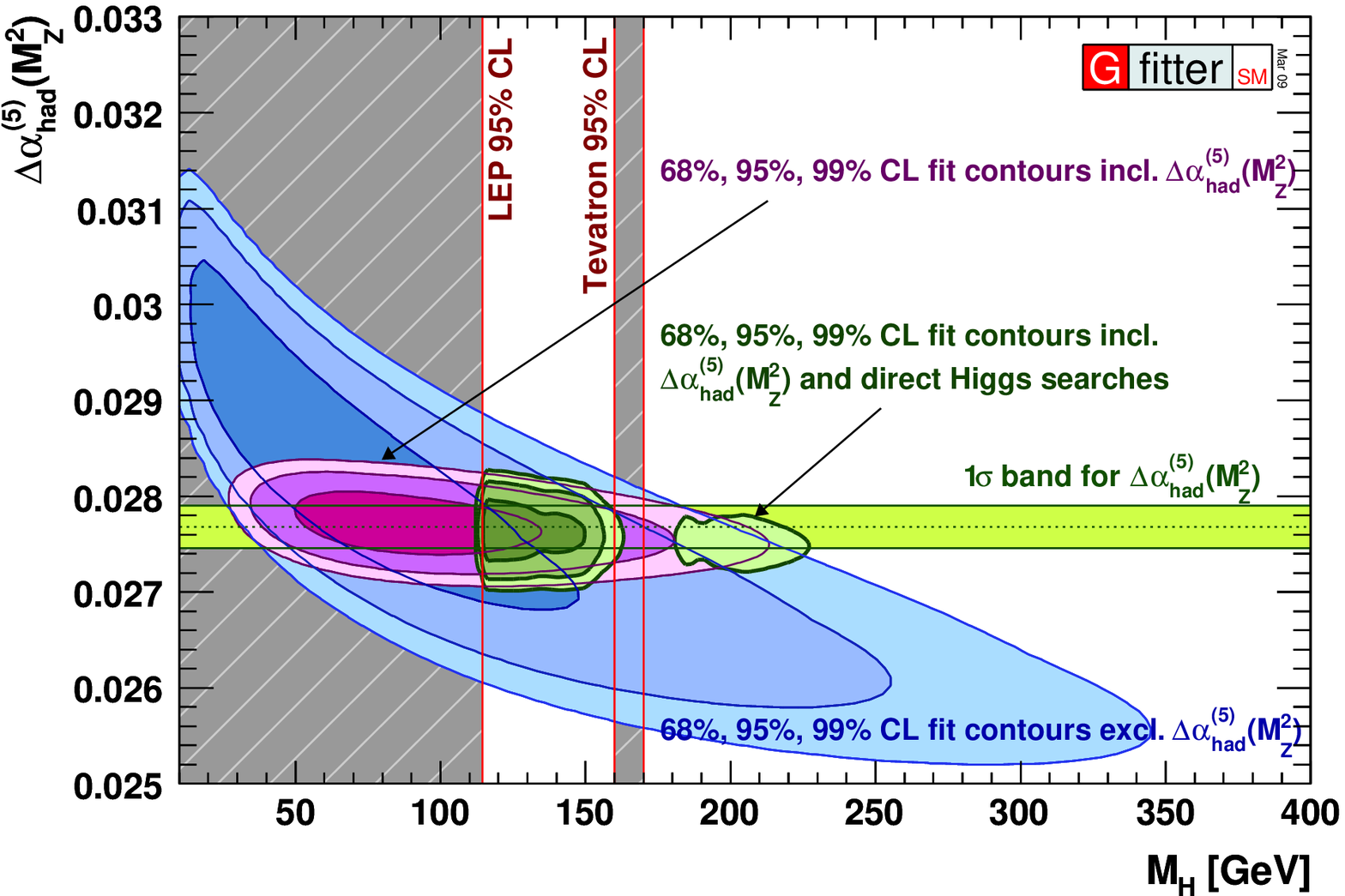}
    \includegraphics[width=\thisSize]{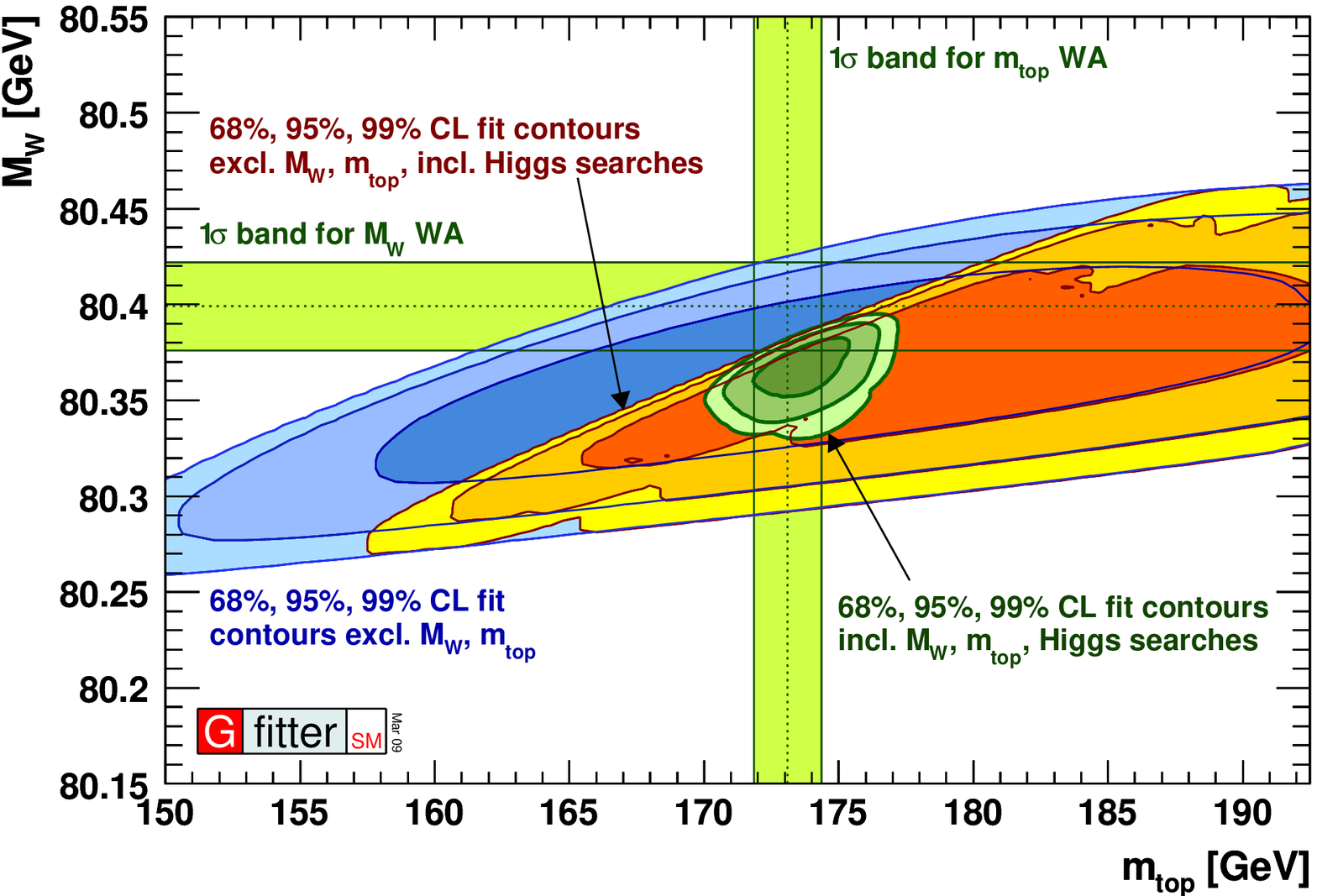}
  \end{center}
  \vspace{-0.3cm}
  \caption{Contours of 68\%, 95\% and 99\%~CL obtained from scans of fits with fixed 
           variable pairs $\mt$ vs. $M_H$ (top), $\Dalphahad$ vs. $M_H$ (middle), and 
           $M_W$ vs. $\mt$ (bottom). The conditions of the various fits shown are indicated 
           on the plots. The horizontal bands depict the $1\sigma$ regions of the current 
           world average measurements (or phenomenological determination in case of 
           $\Dalphahad$). }
\label{fig:2dplots}
\end{figure}

Global fits are performed in two versions: the {\em standard (``blue-band'') fit} makes
use of all the available information except for results from direct Higgs boson searches;
the {\it complete fit} uses also the constraints from the Higgs searches at LEP 
and Tevatron. The free fit parameters are $M_Z$, $M_H$, $m_t$, $m_c$, $m_b$, 
$\Dalphahad$, and \asZ, where only the latter parameter is fully unconstrained (apart from 
$M_H$ in the standard fit). Theoretical uncertainties due to missing higher order perturbative 
corrections in the predictions of $M_W$ and $\sinfeff$ and in the electroweak form factors 
$\rZ{f}$ and $\kZ{f}$ are included in the fit by means of scale parameters according to the 
\Rfit prescription~\cite{ckmfitter}. The relevant input parameters and fit results are 
summarised in Table~\ref{tab:results}, and discussed below.
\begin{table}[p]
\setlength{\tabcolsep}{0.0pc}
{\small
\begin{tabular*}{\textwidth}{@{\extracolsep{\fill}}lccccc} 
\hline\noalign{\smallskip}
& & Free & \multic{2}{c}{Results from global EW fits:} & \multic{1}{c}{\em Complete fit w/o}   \\[-0.1cm]
\rs{Parameter} & \rs{Input value} & in fit & \multic{1}{c}{\em Standard fit} & \multic{1}{c}{\em Complete fit} & \multic{1}{c}{\em exp. input in line} \\
\noalign{\smallskip}\hline\noalign{\smallskip}
$M_{Z}$ {\ft [GeV]} &  $91.1875\pm0.0021$ & yes &  $91.1874\pm0.0021$ &  $91.1876\pm0.0021$ &  $91.1974^{\,+0.0191}_{\,-0.0159}$\\
$\Gamma_{Z}$ {\ft [GeV]} &  $2.4952\pm0.0023$ & -- &  $2.4960\pm0.0015$ &  $2.4956\pm0.0015$ &  $2.4952^{\,+0.0017}_{\,-0.0016}$\\
$\sigma_{\rm had}^{0}$ {\ft [nb]} &  $41.540\pm0.037$ & -- &  $41.478\pm0.014$ &  $41.478\pm0.014$ &  $41.469\pm0.015$\\
$R^{0}_{\l}$ &  $20.767\pm0.025$ & -- &  $20.742\pm0.018$ &  $20.741\pm0.018$ &  $20.717\pm0.027$\\
$A_{\rm FB}^{0,\l}$ &  $0.0171\pm0.0010$ & -- &  $0.01638\pm0.0002$ &  $0.01624\pm0.0002$ &  $0.01617^{\,+0.0002}_{\,-0.0001}$\\
 $A_\ell$ $^{(\star)}$  & $0.1499\pm0.0018$ & --  & $0.1478\pm0.0010$ & $0.1472^{+0.0009}_{-0.0008}$ & --\\
$A_{c}$ &  $0.670\pm0.027$ & -- &  $0.6682^{\,+0.00045}_{\,-0.00044}$ &  $0.6679^{\,+0.00042}_{\,-0.00036}$ &  $0.6679^{\,+0.00041}_{\,-0.00036}$\\
$A_{b}$ &  $0.923\pm0.020$ & -- &  $0.93469\pm0.00010$ &  $0.93463^{\,+0.00007}_{\,-0.00008}$ &  $0.93463^{\,+0.00007}_{\,-0.00008}$\\
$A_{\rm FB}^{0,c}$ &  $0.0707\pm0.0035$ & -- &  $0.0741^{\,+0.0006}_{\,-0.0005}$ &  $0.0737\pm0.0005$ &  $0.0737\pm0.0005$\\
$A_{\rm FB}^{0,b}$ &  $0.0992\pm0.0016$ & -- &  $0.1036\pm0.0007$ &  $0.1032^{\,+0.0007}_{\,-0.0006}$ &  $0.1037^{\,+0.0004}_{\,-0.0005}$\\
$R^{0}_{c}$ &  $0.1721\pm0.0030$ & -- &  $0.17225\pm0.00006$ &  $0.17225\pm0.00006$ &  $0.17225\pm0.00006$\\
$R^{0}_{b}$ &  $0.21629\pm0.00066$ & -- &  $0.21578\pm0.00005$ &  $0.21577\pm0.00005$ &  $0.21577\pm0.00005$\\
$\sinleff(Q_{\rm FB})$ &  $0.2324\pm0.0012$ & -- &  $0.23142\pm0.00013$ &  $0.23151^{\,+0.00010}_{\,-0.00012}$ &  $0.23149^{\,+0.00013}_{\,-0.00010}$\\
\noalign{\smallskip}\hline\noalign{\smallskip}
$M_{H}$ {\ft [GeV]} $^{(\circ)}$ & Likelihood ratios & yes & $ 83^{+ 30[+ 75]}_{- 23[- 41]}$ & $116^{+ 15.6[+36.5]}_{-1.3[-  2.2]}$ & $ 83^{+ 30[+ 75]}_{- 23[- 41]}$\\
\noalign{\smallskip}\hline\noalign{\smallskip}
$M_{W}$ {\ft [GeV]} &  $80.399\pm0.023$ & -- &  $80.384^{\,+0.014}_{\,-0.015}$ &  $80.371^{\,+0.008}_{\,-0.011}$ &  $80.361^{\,+0.013}_{\,-0.012}$\\
$\Gamma_{W}$ {\ft [GeV]} &  $2.098\pm0.048$ & -- &  $2.092^{\,+0.001}_{\,-0.002}$ &  $2.092\pm0.001$ &  $2.092\pm0.001$\\
\noalign{\smallskip}\hline\noalign{\smallskip}
$\mc$ {\ft [GeV]} &  $1.25\pm0.09$ & yes &  $1.25\pm0.09$ &  $1.25\pm0.09$ & -- \\
$\mb$ {\ft [GeV]} &  $4.20\pm0.07$ & yes &  $4.20\pm0.07$ &  $4.20\pm0.07$ & -- \\
$m_{t}$ {\ft [GeV]} &  $173.1\pm1.3$ & yes &  $173.2\pm1.2$ &  $173.6\pm1.2$ &  $179.5^{\,+8.8}_{\,-5.2}$\\
$\dalphaHadMZ$ $^{(\dag\bigtriangleup)}$ &  $2767\pm  22$ & yes & $2772\pm  22$ & $2764^{+  22}_{-  21}$ & $2733^{+  57}_{-  63}$\\
$\alpha_{s}(M_{Z}^{2})$ & -- & yes &  $0.1192^{\,+0.0028}_{\,-0.0027}$ &  $0.1193\pm0.0028$ &  $0.1193\pm0.0028$\\
\noalign{\smallskip}\hline\noalign{\smallskip}
$\deltatheo M_W$ {\ft [MeV]}  & $[-4,4]_{\rm theo}$ & yes  & $4$ & $4$ & -- \\
$\deltatheo \sinleff$ $^{(\dag)}$  & $[-4.7,4.7]_{\rm theo}$ & yes  & $4.7$ & $0.8$ & -- \\
$\deltatheo \rZ{f}$ $^{(\dag)}$ &  $[-2,2]_{\rm theo}$ & yes  & $2$ & $2$ & -- \\
$\deltatheo \kZ{f}$ $^{(\dag)}$ &  $[-2,2]_{\rm theo}$ & yes  & $2$ & $2$ & -- \\
\noalign{\smallskip}\hline
\noalign{\smallskip}
\end{tabular*}
{\ft
$^{(\star)}$Average of LEP and SLD.
$^{(\circ)}$In brackets the $2\sigma$. 
$^{(\dag)}$In units of $10^{-5}$.
$^{(\bigtriangleup)}$Rescaled due to $\alpha_s$ dependency.
}}
\caption[.]{Input values and fit results for parameters of the global electroweak fit. The first and 
         second columns list respectively the observables/parameters used in the fit, and their 
         experimental values or phenomenological estimates (see text for references). 
         The subscript ``theo'' labels theoretical error ranges. 
         The third column indicates whether a parameter is floating in the fit.
         The fourth (fifth) column quotes the results of the {\em standard} ({\em complete}) {\em fit} 
         not including (including) the constraints from the direct Higgs searches at LEP and
         Tevatron in the fit. 
         In case of floating parameters the fit results are directly given,
         while for observables, the central values and errors are obtained by individual profile 
         likelihood scans. The errors are derived from the $\Delta\chi^2$ profile
         using a Gaussian approximation.
         The last column gives the fit results for each parameter without using the corresponding 
         experimental constraint in the fit (indirect determination). 
}         
\label{tab:results}
\end{table}
\begin{itemize}

\item The minimum $\chi^2$ of the standard (complete) fit amounts to 16.4 (17.9) for 13
      (14) degrees of freedom. The corresponding p-value for wrongly rejecting the SM
      from the result of the complete fit, obtained with pseudo-MC samples, is
      $0.20\pm0.01_{-0.02}$, where the first error is statistical and the second is the 
      difference obtained when fixing or varying the theoretical parameters. We also 
      notice that none of the pull values after fit convergence exceeds $3\sigma$.

\item The correlation coefficients between $M_H$ on one hand, and $m_t$, $\Dalphahad$, and $M_W$
      on the other are 0.31, $-0.40$, and $-0.54$, respectively. They are small for all other
      floating fit parameters. In particular, the small correlation with \asZ allows for 
      an independent determination of this quantity, not affected by the unknown Higgs 
      properties. 

\item Some input observables, such as $M_Z$, are much better known than it is required for the 
      fit (the experimental precision exceeds the fit sensitivity by a factor of 10), so that 
      they could have been fixed in the fit without significant change in the results. Others, 
      such as $\Gamma_W$, are not well enough known to impact the fit (the fit 
      sensitivity exceeds the measurement precision by a factor of almost 50). 
      And finally, observables such as $M_W$ and $\mt$ are driving the fit precision.
      Here is where the experimental effort must concentrate on. 

\item We find $M_H=83^{+30}_{-23}\gev$ (standard fit) and $M_H=116^{+16}_{-1.3}\gev$ (complete fit)
      with the $2\sigma$ intervals $[42, 158]\gev$ and $[114, 153]\gev$, respectively (\cf
      top and middle plots in Fig.~\ref{fig:mhiggs}). At $3\sigma$ the complete fit 
      still allows Higgs masses between 180 and 227\gev, which the Tevatron 
      experiments should have the sensitivity to exclude soon. 
      Figure~\ref{fig:2dplots} shows the 68\%, 95\% and 99\% $\CL$ contours for the variable pairs 
      $\mt$ vs. $M_H$ (top plot) and \dalphaHadMZ vs. $M_H$ (middle), exhibiting the largest 
      correlations in the fits.  
      The contours are derived from the \DeltaChi values found in the profile scans using  
      ${\rm Prob}(\DeltaChi,2)$. Three sets of fits are shown in these plots: the largest/blue 
      (narrower/purple) allowed regions are derived from the standard fit excluding (including) 
      the measured values (indicated by shaded/light green horizontal bands) for respectively 
      $\mt$ and \dalphaHadMZ in the fits. The correlations seen in these plots are approximately 
      linear for $\ln M_H$. The third set of fits, providing the narrowest constraints, uses 
      the complete fit, \ie, including in addition to all available measurements the direct 
      Higgs searches. 

\item There is a well-known 
      tension between the $M_H$ results for the most sensitive observables. For 
      $A_\ell({\rm LEP})$, $A_\ell({\rm SLD})$, $A^{0,b}_{\rm FB}$ and $M_W$ we find 
      respectively $M_H=104^{+148}_{-64}$, $26^{+25}_{-16}$, $371^{+295}_{-166}$, and 
      $42^{+56}_{-22}$ (all values in \gev). Evaluating with pseudo-MC experiments, taking 
      into account all known correlations, the probability to observe a value of 
      $\Delta\chi^2 = 8.0$ when removing the least compatible of the four measurements 
      from the fit, gives $1.4\%$ corresponding to an equivalent of $2.5\sigma$.

\item Without using the direct \mt measurement, the standard and complete fits determine the 
      top mass to be $177.2^{+10.5}_{-7.8}\gev$ and $179.5^{+8.8}_{-5.2}\gev$, respectively,
      where the latter result is $1.2\sigma$ away from the experimental value requiring
      a smaller $M_H$, which is excluded by LEP, or a smaller $M_W$ value (see next bullet).
      It is noticeable that the standard fit without using the measured top-mass gives
      $M_H=116^{+184}_{-61}\gev$ with a central value equal to the complete fit (though with
      largely inflated errors). 

\item One can also indirectly determine $M_W$ from the fit without using the input from the 
      direct measurements. The results from the standard and complete fits read 
      $M_W=80.374^{+0.019}_{-0.038}\gev$ and $80.361^{+0.013}_{-0.012}\gev$, respectively,
      requiring a smaller value when including the $M_H$ constraints.
      The bottom plot in Fig.~\ref{fig:2dplots} compares the direct measurements of \mt 
      and $M_W$, shown by the shaded/green $1\sigma$ bands, with the 
      68\%,  95\% and 99\%~CL constraints obtained with three fit scenarios. The largest/blue
      (narrowest/green) allowed regions are the result of the standard fit (complete fit) 
      excluding (including) the measured values of $M_W$ and $\mt$. The results of the complete fit
      excluding the measured values are illustrated by the narrower/yellow allowed region. The
      allowed regions of the indirect determination is significantly reduced with the insertion
      of the direct Higgs searches. 

\item Among the most important outcomes of the fit is the 3NLO~\cite{Baikov:2008jh} precision 
      determination of $\asZ$ obtained mainly by the parameter $R_\ell^0$. One finds
      $\asZ=0.1193\pm0.0028\pm0.0001$, where the first error is experimental and the second 
      due to the truncation of the perturbative series. It includes variations of the 
      renormalisation scale between $0.6\,M_Z<\mu<1.3\,M_Z$~\cite{Davier:2008sk}, 
      of massless terms of order $\as^5(M_Z)$ and higher, and of quadratic massive terms of 
      order and beyond $\as^4(M_Z)$. The result is in excellent agreement with the 
      precise 3NLO determination from hadronic $\tau$ decays, which, evolved to 
      the $Z$-mass scale, reads $\asZ=0.1212\pm0.0005\pm0.0009$\cite{Davier:2008sk}, 
      dominated by theoretical uncertainties.\footnote
      {
         Only partly contained in the theoretical error are systematic differences arising 
         from the computation of the contour integral, denoted as fixed-order perturbation
         theory (FOPT) and contour-improved fixed-order perturbation perturbation theory
         (CIPT), respectively. The value cited here uses CIPT. The differences between FOPT
         and CIPT are discussed in Refs.~\cite{Baikov:2008jh,Davier:2008sk,beneke,menke,maltman}.
      }
      These two measurements represent the best current 
      test of the asymptotic freedom property of QCD (\cf Fig.~\ref{fig:alphas}).

\end{itemize}
\begin{figure}[t]
\centerline{\includegraphics[width=0.730\columnwidth]{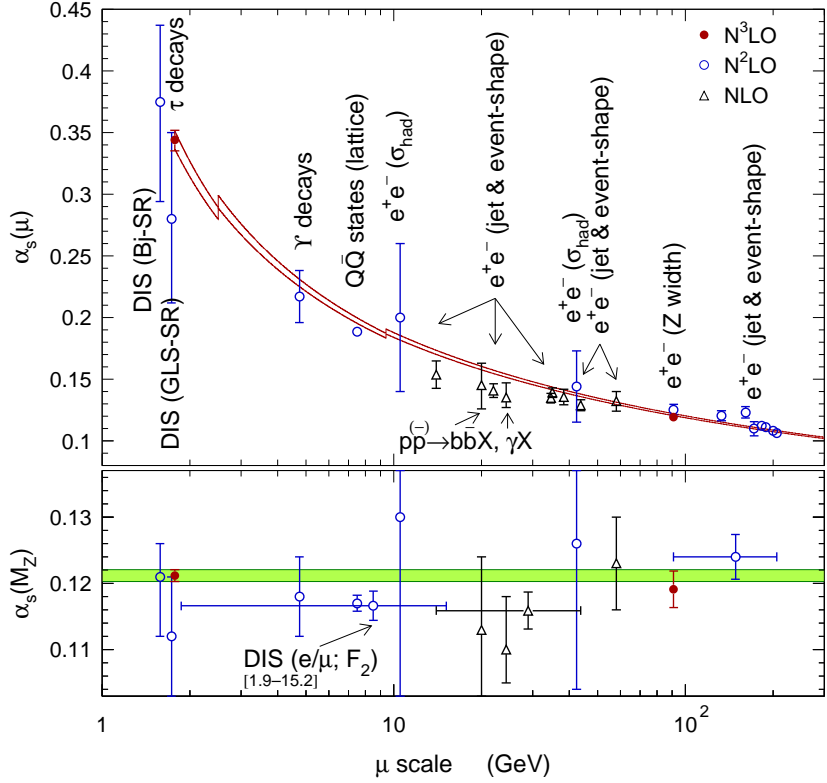}}
\vspace{-0.5cm}
\caption{Evolution of \asTau to higher scales $\mu$ using 
         the four-loop RGE and the three-loop matching conditions applied at the
         heavy quark-pair thresholds (hence the discontinuities at $2\overline m_c$ 
         and $2\overline m_b$). The evolution is compared with measurements at different
         scales (from Ref.~\cite{bethke2006} and including newer measurements) 
         covering more than two orders of magnitude. The bottom part shows the corresponding 
         $\as$ values evolved to $M_Z$. The shaded band displays the $\tau$ decay 
         result within errors. Only the $\tau$ and $Z$-scale measurements have 3NLO 
         theoretical accuracy. The figure is taken from Ref.~\cite{Davier:2008sk}.}
\label{fig:alphas}
\end{figure}

\section{Future prospects}

\begin{figure}[t]
  \centering
  \begin{center}
     \includegraphics[width=0.685\textwidth]{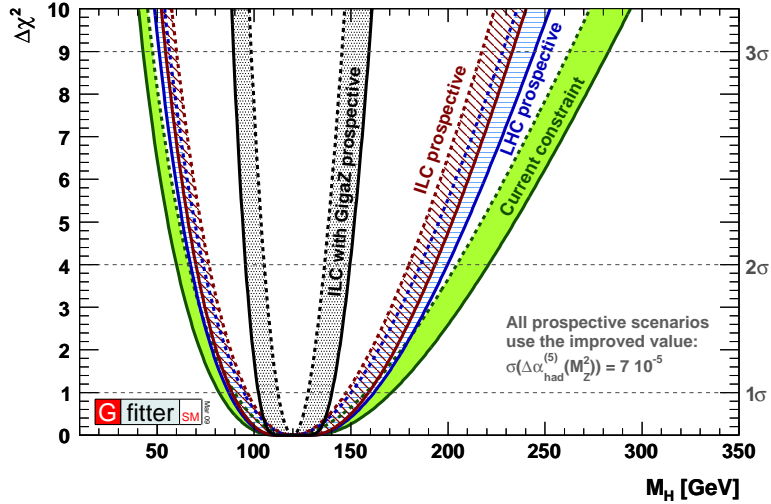}
  \end{center}
  \vspace{-0.3cm}
  \caption[.]{
            Constraints on $M_H$ obtained for the four future 
            scenarios discussed in the text.
            An improvement of $\sigma(\dalphaHadMZ)=7\cdot 10^{-5}$ is assumed for all 
            prospective curves. The shaded bands indicate the contributions from theoretical 
            uncertainties in the EW theory. 
          }
  \label{fig:future}
\end{figure}
Several improved measurements are expected from the LHC~\cite{atlastdr}. The Higgs boson 
should be discovered leaving the SM without an unmeasured parameter.\footnote
{ 
  Excluding here the massive neutrino sector, requiring at least nine additional 
  parameters, which are however irrelevant for the results discussed here. 
}
The focus of the global SM fit would then move from parameter estimation to the analysis 
of the goodness-of-fit with the goal to uncover inconsistencies between the model and 
the data, indicating the presence of new physics. Because the Higgs-boson mass enters 
only logarithmically in the loop corrections, a precision measurement is not required 
for this purpose. With the LHC the uncertainty on the $W$-boson and the top-quark masses 
should shrink to $15\mev$ and $1\gev$, respectively. The fit constraint on $M_H$ for a 
hypothetical 120\gev Higgs boson would improve from currently $120^{+50}_{-40}\gev$ to 
$120^{+45}_{-35}\gev$ (assuming unchanged theoretical errors).

At the ILC, a significant increase in the top mass precision to an error of at least
$0.2\gev$ from a threshold scan is expected, providing a Higgs mass constraint of 
$120^{+42}_{-33}\gev$. If in the meantime the prediction of $\dalphaHadMZ$ has been 
improved to an accuracy of, say, $7\cdot 10^{-5}$, by virtue of more accurate hadronic 
cross section data at low and intermediate energies, one could achieve $M_H=120^{+39}_{-31}\gev$.

Running the ILC at lower energy with polarised beams (Giga-Z), the $W$ and top masses could 
be determined to better than $6\mev$ and $0.1\gev$, respectively. Moreover, the weak 
mixing angle is expected to be measured to a precision of $1.3\cdot10^{-5}$~\cite{Djouadi}, 
and $R_\ell^0$ to 0.004, resulting in an unprecedented precision determination of $\asZ$ 
with an error of 0.0006, \ie, a factor of 3 improvement over the current value.
Owing to the small theoretical error at 3NLO such a precision could be fully 
exploited~\cite{Baikov:2008jh}. The improvements on the prediction of the Higgs mass 
are dramatic for Giga-Z where we find for a 120\gev Higgs fit errors of about 19\gev, 
assuming the improved $\Dalphahad$. Such a constraint on $M_H$ will be significantly 
affected by from theoretical errors (without theory errors one could constrain the 
Higgs mass to $8\gev$), requiring improved electroweak calculations. 

The $M_H$ scans obtained for the four scenarios, assuming the improved \dalphaHadMZ 
precision to be applicable for all future scenarios, are shown in Fig.~\ref{fig:future}. 
The shaded bands indicate the effects of the current theoretical uncertainties.
The theoretical errors treated with the \Rfit scheme are recognised by the broad 
plateaus around the $\Delta\chi^2$ minimum.

In case of the discovery of a light Higgs, and a precise mass measurement (expected 
to be 0.1\% or better for $H\to\gamma\gamma$), the $W$ 
boson mass could be predicted with $13\mev$ error, of which $5\mev$ is (currently) 
theoretical. With the new machines, new precision measurements would enter the fit, 
namely the two-fermion cross section at higher energies and the electroweak triple 
gauge boson couplings, which are sensitive to models beyond the SM.
Most importantly however, both machines are directly sensitive to new phenomena
and thus either provide additional constraints on fits of new physics models or, if the 
searches are successful, may completely alter our picture of the terascale physics. 
The SM will then require extensions, the new parameters of which must be determined by a 
global fit, whose goodness must also be probed. 

\section{The fate of the Standard Model}

\begin{figure}[t]
  \begin{center}
  \includegraphics[width=0.685\textwidth]{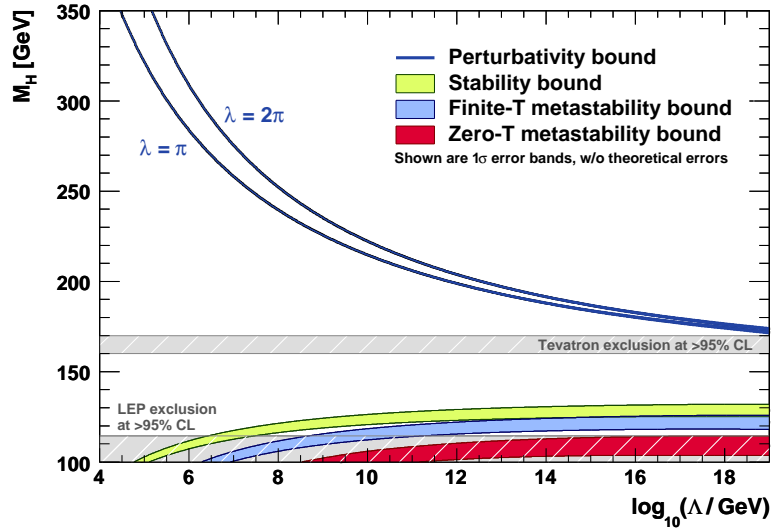}
  \end{center}
  \vspace{-0.3cm}
\caption[.]{The scale $\Lambda$ at which the two-loop RGEs drive the quartic SM Higgs 
         coupling non-perturbative, and the scale $\Lambda$ at which the RGEs 
         create an instability in the electroweak vacuum ($\lambda < 0$). 
         The width of the bands indicates the errors 
         induced by the uncertainties in the top mass and in $\as$ (added quadratically).         
         The perturbativity upper bound (sometimes referred to as 
         ``triviality'' bound) is given for $\lambda = \pi$ 
         (lower bold line) and $\lambda =2\pi$ 
         (upper bold line). Their difference indicates the size of the
         theoretical uncertainty in this bound. The absolute 
         vacuum stability bound is displayed by the light shaded (green) band,
         while the less restrictive finite-temperature and zero-temperature 
         metastability bounds are medium (blue) and dark shaded (red), respectively.
         The theoretical uncertainties in these bounds have been ignored in 
         the plot for the purpose of clarity. 
         The grey hatched areas indicate the LEP~\cite{Higgs-LEP} and 
         Tevatron~\cite{Higgs-Tev} exclusion domains. The figure is taken 
         from Ref.~\cite{fate}.}
\label{fig:bounds}
\end{figure}
The Higgs sector of the SM must steer a narrow course between two problematic situations 
if it is to survive up to the reduced Planck scale $M_P \sim 2\cdot 10^{18}\gev$.
If $M_H$ is large enough, the renormalisation-group equations (RGEs) of the SM drive 
the Higgs self-coupling into the nonperturbative regime at some scale $\Lambda < M_P$, 
entailing either new non-perturbative physics at a scale $\sim \Lambda$, or new physics
at some scale $< \Lambda$ that prevents the Higgs self-coupling from becoming nonperturbative.
This is shown as the upper pair of bold (blue) lines in Fig.~\ref{fig:bounds}~\cite{fate}. 
On the other hand, if $M_H$ is small enough, the RGEs drive the Higgs self-coupling 
to a negative value at some Higgs field value $\Lambda < M_P$, in which case the 
electroweak vacuum is only a local minimum and there is a new, deep and potentially 
dangerous minimum at scales $>\Lambda$. The electroweak vacuum can 
become unstable against collapse because of zero-temperature  
or thermal tunneling during the evolution of the universe into that deeper 
new vacuum with Higgs vacuum expectation value $> \Lambda$, unless there is new physics 
at some scale $< \Lambda$ that prevents the appearance of that vacuum. 
This lower bound is shown with its uncertainties by the light shaded (green) band in
Fig.~\ref{fig:bounds}.  Below this stability bound, there is a metastability region, 
where the total quantum tunneling probability throughout the period of the history 
of the Universe is small enough so that the electroweak
vacuum has a lifetime longer than the age of the Universe for decay via either
zero-temperature quantum fluctuations (region above the dark shaded (red)
band in Fig.~\ref{fig:bounds}) or thermal fluctuations (region above the medium
shaded (blue) band). Uncertainties in these bounds stem from top-mass and \as
dependencies as well as theoretical errors mainly due to missing higher 
order RGE corrections. At $\Lambda=M_P$, the bounds read~\cite{fate}: 
$M_H<(175\;(173)\pm0.8\pm0.1)\gev$ (nonperturbative bound for $\lambda(M_P) = 2\pi\;(\pi)$),
$M_H>(108.9\pm5.3\pm3.0)\gev$ (zero-temperature metastability),
$M_H>(122.0\pm3.7\pm3.0)\gev$ (thermal metastability), and
$M_H>(128.6\pm3.4\pm1.0)\gev$ (absolute vacuum stability).
Between these bounds there is a range of intermediate values of $M_H$ for which 
the SM could survive up to the Planck scale. 

The bounds can be convolved with the $M_H$ constraints from the global 
electroweak fit including the direct Higgs searches~\cite{fate}. In particular
the Tevatron data~\cite{Higgs-Tev} increase the exclusion of the nonperturbative 
scenario from 95.7\% to 99.1\%~CL, if $\lambda=2\pi$ is taken to be the nonperturbative 
threshold. The collapse scenario is ruled out for most of the parameter 
region by the direct Higgs searches at LEP~\cite{Higgs-LEP}, though a small region 
is still compatible with the limit so that no significant  exclusion CL can 
be set (we find a p-value of 0.40 for it being compatible with the LEP result). 

\begin{figure}[t]
\begin{center}
     \includegraphics[width=0.497\textwidth]{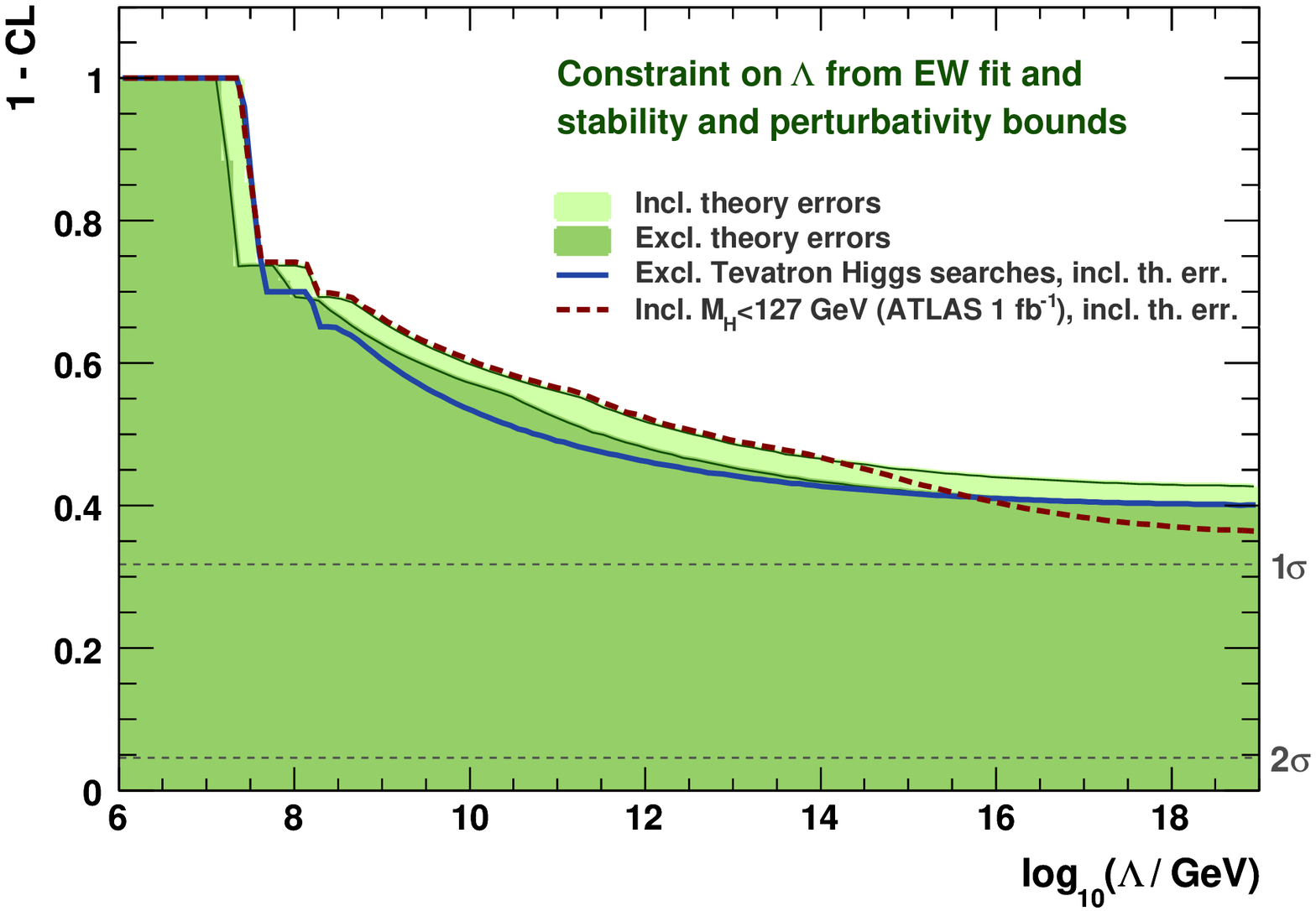}
     \includegraphics[width=0.497\textwidth]{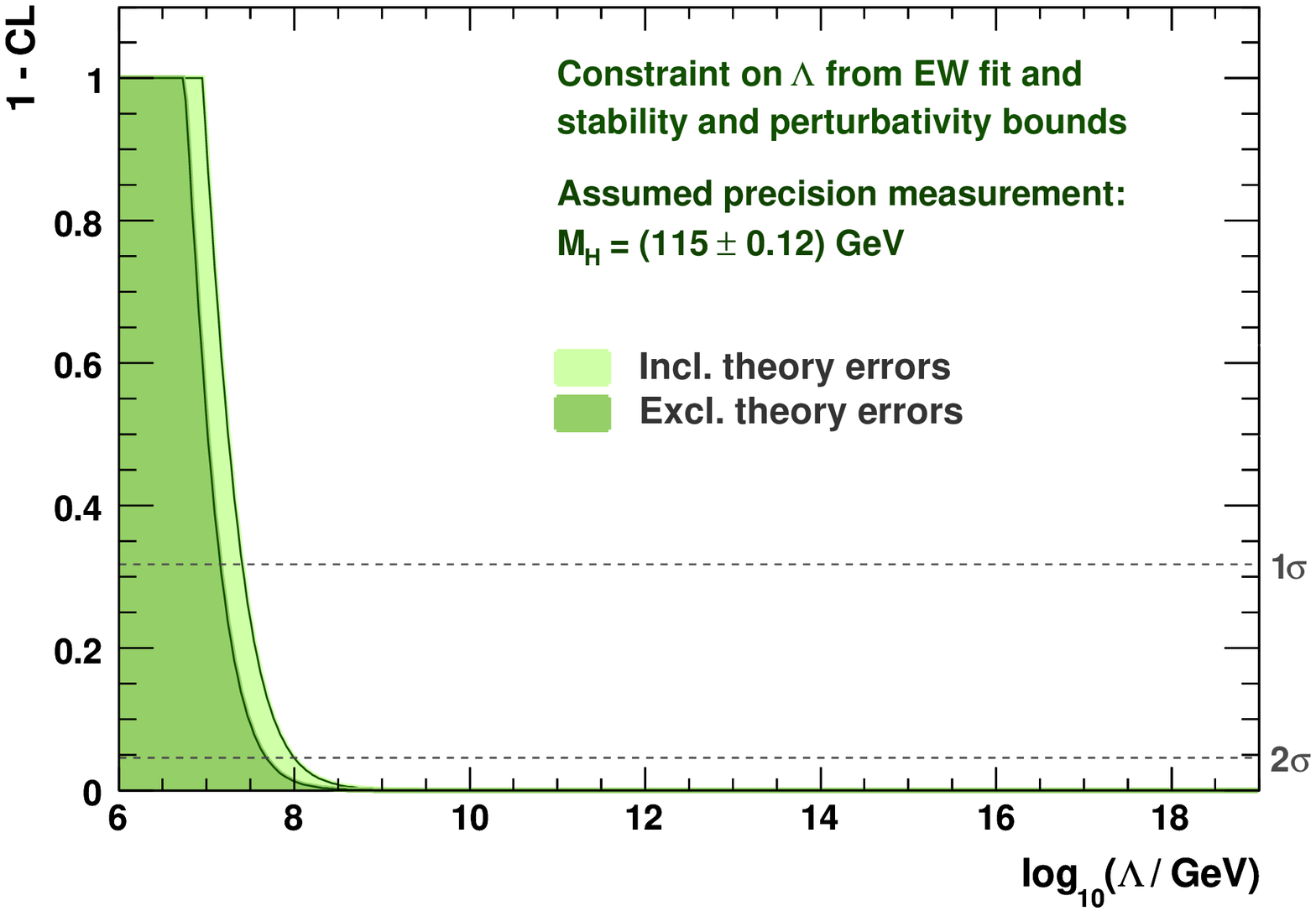}
\end{center}
  \vspace{-0.3cm}
\caption{\underline{Left:} constraint on $\Lambda$, expressed as 1\,$-$\,CL, from 
         the global electroweak fit and the requirement of absolute vacuum stability 
         and perturbativity. Shown are fits
         with (light shading) and without (dark shading) taking into account 
         the theoretical uncertainty in the stability bound. The bold solid (blue)
         line shows the effect of removing the Tevatron Higgs searches from the 
         global fit. The dashed (red) line shows the effect of a hypothetical upper 
         bound $M_H < 127\gev$ at 95\%~CL, as might be obtained with early data at 
         the LHC.
         \underline{Right:} assuming in addition the discovery of a light Higgs
         with precisely measured mass of 115\gev. Also included in the fits of 
         this plot are improved errors for the top and $W$ masses, as anticipated 
         for the LHC.}
\label{fig:lam_mh}
\end{figure}
Requiring absolute vacuum stability and perturbativity until $\Lambda=M_P$, and using
all the available constraints on $M_H$, one can derive confidence levels for the 
maximum allowed scale $\Lambda$ before new physics must come in to stabilise 
the Higgs potential. Figure~\ref{fig:lam_mh} (left) shows $1-\CL$ versus $\Lambda$ 
for various cases: with (light shaded) and without (dark shaded) the theoretical 
uncertainty in the stability bound, including and excluding (solid line) the results
from the Tevatron Higgs boson searches, and assuming a hypothetical unsuccessful early
Higgs search at one of the high-$p_T$ LHC experiments (represented here by
ATLAS), for an integrated luminosity of approximately $1\:{\rm fb}^{-1}$
at 14\tev centre-of-mass energy, that should have sufficient sensitivity
to exclude $M_H > 127\gev$ at 95\%~CL~\cite{atlas-cscbook} (dashed line).
No constraint on $\Lambda$ that would reach or exceed 68\%~CL can be derived from
the present data, nor from the prospective incremental improvement in the
Higgs constraint that might come from the Tevatron or the early
running of the LHC. If, however, there were a Higgs
discovery with a mass determined to be $M_H=115\gev$ after years of successful LHC 
operation, one would obtain the constraints on $\Lambda$ plotted in the right plot 
of Fig.~\ref{fig:lam_mh}. 
The 95\%~CL upper limits on the cut-off scale, obtained including theoretical errors, 
would read $\log_{10}(\Lambda/\gev)<10.4$ and $8.0$, respectively (including an almost 
half an order of magnitude theoretical uncertainty). In this case, one would obtain 
an upper limit on the absolute stability of the SM that would be comparable
with the scale suggested by the seesaw model for the light neutrino
masses. The p-value of the $M_H=115\gev$ scenario for the survival of the 
SM up to $M_P$ is as small as the occurrence of a $5.3\sigma$ fluctuation.

\section{Concluding remarks}

The efforts by many
to develop with the Gfitter package a modern tool for model testing in High-Energy 
Physics has found its first application in a reimplementation of the global electroweak 
fit leading, together with the direct Higgs boson searches, to an exclusion of Standard 
Model Higgs masses above 153\gev at 95\% confidence level. Exploiting new theoretical 
developments, the fit also provides the, as of today, theoretically most robust 
determination of \as. Gfitter allows us to quantify the narrow passage for the Standard 
Model to survive all the way up to the Planck scale, between such catastrophic scenarios 
as nonperturbative blow-up or a collapsing electroweak vacuum~\cite{fate}. Similar to the 
hierarchy problem, these scenarios manifest the instability of the Higgs field under 
radiative corrections. 

All results presented in this review are obtained in the framework of the minimal Standard 
Model. Extensions of the Higgs sector~\cite{grojean} may evade the constraints form the 
electroweak precision data. The effects of these extensions on the gauge vector-boson 
self-energy graphs, known as oblique corrections, are known for most of the models and must 
be continuously confronted with the newest experimental data. Physics beyond the Standard
Model should alter the high-scale behaviour of the Higgs potential, thus possibly rendering 
the upper bound on the Higgs mass derived from nonperturbativity irrelevant. Due to 
these arguments, and the strong theoretical grounds for new physics at the TeV scale, 
experimental Higgs searches cannot rely on the limits obtained with fits within the 
Standard Model, but must continue to explore all the sensitive phase space not yet 
excluded by direct searches. 

The imminent goals for the Gfitter 
group are two-fold: ($i$) maintain the Standard Model package in line with experimental 
and theoretical progress, and continuously improve the Gfitter core package and the fit 
efficiency, ($ii$) extend it by plugging in new physics models. Examples for 
analyses beyond the Standard Model in Gfitter are the Type-II Two-Higgs-Doublet 
model~\cite{baak,gfitter} and oblique parameter fits~\cite{goebel}. The latter 
analysis will be diversified and, among others, augmented by Supersymmetry.

\end{document}

%% file: definitions.tex
\newcommand{\bbar}  {\ensuremath{\overline b}\xspace}

\mathchardef\Upsilon="7107
\def\Y#1S{\ensuremath{\Upsilon{(#1S)}}\xspace}

\newcommand{\mtau}{\ensuremath{m_\tau}\xspace}
\newcommand{\mt}{\ensuremath{m_{t}}\xspace}
\newcommand{\as}{\ensuremath{\alpha_{\scriptscriptstyle S}}\xspace}

\newcommand{\asTau}{\ensuremath{\as(\mtau^2)}\xspace}
\newcommand{\asZ}{\ensuremath{\as(M_Z^2)}\xspace}

\renewcommand\l{\ell}

\newcommand{\GF}{{\ensuremath{G_{\scriptscriptstyle F}}}\xspace}

\newcommand{\Kbar    }{\kern 0.2em\overline{\kern -0.2em K}{}\xspace}

\newcommand{\Kz      }{\ensuremath{K^0}\xspace}
\newcommand{\Kzb     }{\ensuremath{\Kbar^0}\xspace}
\newcommand{\KzKzb   }{\ensuremath{\Kz \kern -0.16em \Kzb}\xspace}
\newcommand{\Kp      }{\ensuremath{K^+}\xspace}
\newcommand{\Km      }{\ensuremath{K^-}\xspace}

\newcommand{\KpKm    }{\ensuremath{\Kp \kern -0.16em \Km}\xspace}

\newcommand\Dbar    {\kern 0.18em\overline{\kern -0.18em D}{}\xspace}

\newcommand\Bbar    {\kern 0.18em\overline{\kern -0.18em B}{}\xspace}

\newcommand\Bz      {\ensuremath{B^0}\xspace}

\newcommand\Bzb     {\ensuremath{\Bbar^0}\xspace}
\newcommand\Bu      {\ensuremath{B^+}\xspace}
\newcommand\Bub     {\ensuremath{B^-}\xspace}

\newcommand\BpBm    {\ensuremath{\Bu {\kern -0.16em \Bub}}\xspace}
\newcommand\Bs      {\ensuremath{B^0_{s}}\xspace}
\newcommand\Bsb     {\ensuremath{\Bbar^0_{s}}\xspace}

\newcommand\BzBzb   {\ensuremath{\Bz {\kern -0.16em \Bzb}}\xspace}
\newcommand\BszBszb {\ensuremath{\Bs {\kern -0.16em \Bsb}}\xspace}

\newcommand\invfb{\ensuremath{\;{\rm fb}^{-1}}\xspace}

\newcommand\deltatheo{\ensuremath{\delta_{\rm th}}\xspace}

\newcommand\Rfit{{\em R}fit\xspace}

\newcommand{\ft}{\footnotesize}
\newcommand{\multic}{\multicolumn}

\newcommand{\tev}{\ensuremath{\mathrm{\;Te\kern -0.1em V}}\xspace}
\newcommand{\gev}{\ensuremath{\mathrm{\;Ge\kern -0.1em V}}\xspace}
\newcommand{\mev}{\ensuremath{\mathrm{\;Me\kern -0.1em V}}\xspace}
\newcommand{\kev}{\ensuremath{\mathrm{\;ke\kern -0.1em V}}\xspace}
\newcommand{\ev}{\ensuremath{\mathrm{\,e\kern -0.1em V}}\xspace}
\newcommand{\gevc}{\ensuremath{{\mathrm{\,Ge\kern -0.1em V\!/}c}}\xspace}
\newcommand{\mevc}{\ensuremath{{\mathrm{\,Me\kern -0.1em V\!/}c}}\xspace}
\newcommand{\gevcc}{\ensuremath{{\mathrm{\,Ge\kern -0.1em V\!/}c^2}}\xspace}
\newcommand{\mevcc}{\ensuremath{{\mathrm{\,Me\kern -0.1em V\!/}c^2}}\xspace}

\newcommand{\bei}{\begin{itemize}}
\newcommand{\eei}{\end{itemize}}
\newcommand{\beq}{\begin{equation}}
\newcommand{\eeq}{\end{equation}}
\newcommand{\beqn}{\begin{eqnarray}}
\newcommand{\eeqn}{\end{eqnarray}}
\newcommand{\beqns}{\begin{eqnarray*}}
\newcommand{\eeqns}{\end{eqnarray*}}
\newcommand{\bitm}{\begin{itemize}}
\newcommand{\eitm}{\end{itemize}}

\newcommand{\dalphaHadMZ}{\ensuremath{\Delta\alpha_{\rm had}^{(5)}(M_Z^2)}\xspace}

\newcommand\ie{{\it i.e.}\xspace}

\newcommand\cf{{\em cf.}\xspace}
\newcommand{\ea}{{\em et al.}\xspace}
\newcommand\rs{\raisebox{1.5ex}[-1.5ex]}

\def\@citex[#1]#2{\if@filesw\immediate\write\@auxout{\string\citation{#2}}\fi
  \@tempcnta\z@\@tempcntb\m@ne\def\@citea{}\@cite{\@for\@citeb:=#2\do
    {\@ifundefined
       {b@\@citeb}{\@citeo\@tempcntb\m@ne\@citea
        \def\@citea{,\penalty\@m\ }{\bf ?}\@warning
       {Citation `\@citeb' on page \thepage \space undefined}}%
    {\setbox\z@\hbox{\global\@tempcntc0\csname b@\@citeb\endcsname\relax}%
     \ifnum\@tempcntc=\z@ \@citeo\@tempcntb\m@ne
       \@citea\def\@citea{,\penalty\@m}
       \hbox{\csname b@\@citeb\endcsname}%
     \else
      \advance\@tempcntb\@ne
      \ifnum\@tempcntb=\@tempcntc
      \else\advance\@tempcntb\m@ne\@citeo
      \@tempcnta\@tempcntc\@tempcntb\@tempcntc\fi\fi}}\@citeo}{#1}}

\def\@citeo{\ifnum\@tempcnta>\@tempcntb\else\@citea
  \def\@citea{,\penalty\@m}%
  \ifnum\@tempcnta=\@tempcntb\the\@tempcnta\else
   {\advance\@tempcnta\@ne\ifnum\@tempcnta=\@tempcntb \else
\def\@citea{--}\fi
    \advance\@tempcnta\m@ne\the\@tempcnta\@citea\the\@tempcntb}\fi\fi}

\xspace

\newcommand\CL{{\rm CL}}

\newcommand\DeltaChi{\ensuremath{\Delta\chi^2}\xspace}

\newcommand{\Dalphahad}{\dalphaHadMZ}

\newcommand{\seffsf}[1]{\sin\!^2\theta^{#1}_{{\rm eff}}}

\newcommand{\rZ}[1]{\rho^{#1}_{Z}}

\newcommand{\kZ}[1]{\kappa^{#1}_{Z}}

\newcommand{\sinfeff}{\sin\!^2\theta^f_{{\rm eff}}}
\newcommand{\sinleff}{\seffsf{\ell}}

\newcommand{\mc }{\ensuremath{\overline{m}_c}\xspace}

\newcommand{\mb }{\ensuremath{\overline{m}_b  }\xspace}
